\documentclass[prd,tightenlines,nofootinbib,showpacs,preprintnumbers,superscriptaddress, longbibliography, notitlepage]{revtex4-1}

\usepackage{amsfonts,amsmath,amssymb,amsthm,bbm,hyperref}
\usepackage{graphicx}
\usepackage{color}

\newcommand{\be}{\begin{equation}}
\newcommand{\ee}{\end{equation}}
\newcommand{\beq}{\begin{eqnarray}}
\newcommand{\eeq}{\end{eqnarray}}

\begin{document}

\title{Time reversal symmetry in cosmology and the creation of a universe-antiuniverse pair}
\author{Salvador J. Robles-P\'{e}rez}
\affiliation{Estaci\'{o}n Ecol\'{o}gica de Biocosmolog\'{\i}a, Pedro de Alvarado, 14, 06411 Medell\'{\i}n, Spain.}
\affiliation{Departamento de matem\'{a}ticas, IES Miguel Delibes, Miguel Hern\'{a}ndez 2, 28991 Torrej\'{o}n de la Calzada, Spain.}

\date{\today}

\begin{abstract}
The classical evolution of the universe can be seen as a parametrised worldline of the minisuperspace, with the time variable $t$ the parameter that parametrises the worldline. The time reversal symmetry of the field equations implies  that for any positive oriented solution there can be a symmetric negative oriented one that, in terms of the same time variable, represent an expanding and a contracting universe, respectively. However, the choice of the time variable induced by the correct value of the Schrödinger equation in the two universes makes that their physical time variables could be reversely related. In that case, the two universes would be both expanding universes from the point of view of their internal inhabitants, who identify matter with the particles that move in their spacetimes and antimatter with the particles that move in the time reversely symmetric universe. If the assumptions considered are consistent with a realistic scenario of our universe, the creation of a universe-antiuniverse pair might explain two main and related problems in cosmology: the time asymmetry and the primordial matter-antimatter asymmetry of our universe.
\end{abstract}


\maketitle


\section{Introduction}

There is a formal analogy between the evolution of the universe in the minisuperspace and the trajectory of a test particle in a curved spacetime. The former is given, for a homogeneous and isotropic universe, by the solutions of the field equation $a(t)$ and $\vec \varphi(t) = (\varphi_1(t), \ldots, \varphi_n(t))$, where $a$ is the scale factor and $\varphi_i$ are $n$ scalar fields that represent the matter content of the universe. The evolution of the universe can then be seen as a parametrised trajectory in the $n+1$ dimensional space formed by the coordinates $a$ and $\vec \varphi$, which is called the minisuperspace. The trajectory  is the worldline that extremizes the Einstein-Hilbert action, the time variable $t$ is the parameter that parametrises the worldline, and the parametric coordinates along the worldline are the classical solutions, $(a(t), \vec \varphi(t))$ .

From that point of view, the time reversal invariance of the laws of physics translates in the minisuperspace into the invariance that we have in running the worldline in the two possible directions, forward and backward, along the worldline. It is similar to what happens with the trajectory of a test particle in the spacetime. In particle physics, Feynman interpreted the time forward and the time backward solutions of the trajectory of a test particle as the trajectories of the particles and antiparticles of the Dirac's theory \cite{Feynman1949}. In cosmology, we can also assume that the two symmetric solutions may form a universe-antiuniverse pair. In the universe, however, a forward oriented trajectory with respect to the scale factor component, $\dot a >0$, means an increasing value of the scale factor so it represents an expanding universe. Similarly, a backward solution ($\dot a <0$) represents a contracting universe. Therefore, the created pair contains, in terms of the same time variable, a contracting universe and an expanding universe. 

However, the analysis of the emergence of the classical spacetime in quantum cosmology suggests that the time variables of the two universes should be reversely related \cite{RP2017c, RP2018a}, $t_1 = - t_2$. In that case, the matter that propagates in one of the universes can naturally be identified, from the point of view of the symmetric universe, with antimatter, and viceversa. Let us notice that, from the point of view of quantum cosmology, the semiclassical picture of quantum matter fields propagating in a classical background spacetime is an emergent feature that appears, after some decoherence process, in the semiclassical regime \cite{Hartle1990, Kiefer1992}. In that case, we shall see in this paper that in order to obtain the correct value of the Schrödinger equation in the two universes their time variables would be reversely related. Then, the time variables measured by the internal observers in their particle physics experiments, i.e. the time variables that appear in the Schrödinger equation of their physical experiments, would be reversely related and, from that point of view, the matter that propagates in a hypothetical partner universe could naturally de identified with the primordial antimatter that is absent in the observer's universe. From the global point of view of the composite state, then, the apparent asymmetry between matter and antimatter would be restored.

A caveat should be made however on the assumptions of homogeneity and  isotropy  for the initial spacetime manifold, which is in the basis of the time reversal symmetry of the cosmological field equations and it is therefore a condition for the creation of universes in pairs with reversely related time variables. From the point of view of a full quantum theory of gravity it is expected the creation of all kind of universes with all kind of (even exotic) geometries of their spacetimes, so the creation of a homogeneous and isotropic universe should be considered as a particular case, and the consequent scenario of the creation of a universe-antiuniverse pair only as a plausible one. Even though, the scenario might be rather realistic provided that we assume that the fluctuations of the spacetime are relatively small from the very onset of the universe. The observational data suggest that at least from a very remote past our universe essentially looks homogeneous and isotropic, with relatively small inhomogeneities and anisotropies compared with the energy of the homogeneous and isotropic background. Accordingly, we shall assume in this paper that the universe left the Euclidean gravitational vacuum and started inflating from an initial spatial hypersurface, $\Sigma(a_i)$, that is small but large enough to assume that the fluctuations of the spacetime are subdominant, i.e. $a_i \gg l_P$. It is not an unrealistic scenario. For instance, in the Higgs-inflaton scenario \cite{Bezrukov2008, RP2019b}, $a_i \sim V^{-1/2} \propto \xi \, l_P$, where $V$ is the potential of the Higgs in the initial slow roll regime and $\xi \gg 1$ is the strong coupling between the Higgs and gravity. We shall later on comment on the effect that the fluctuations of the spacetime would have on the breaking of the time reversal symmetry.

In this paper we review and gather the main results of previous works \cite{RP2011b, RP2014, RP2017a, RP2017c, RP2018a} and extend the hypothesis  presented in \cite{RP2017e} for the restoration of the primordial matter-antimatter asymmetry to the more general scenario of two homogeneous and isotropic pieces of the spacetime whose time variables are, according to the time reversal symmetry of the Einstein-Hilbert action, reversely related. That will prepare  the arena for more detailed, future developments on the subject. In Sect. \ref{sec2}, we present the analogy between the classical evolution of the universe in the minisuperspace and the trajectory of a test particle in a curved spacetime. It is shown that the time reversal symmetry of the action and the conservation of the total momentum in the process of creation of the universe would imply that the universes should be created in pairs with opposite values of their momenta so that the total momentum is zero. In Sect. \ref{sec3}, it is analysed the correlations between the spacetimes that emerge from the two wave functions that are associated to the opposite values of the momenta conjugated to the scale factor. It turns out that in order to obtain the correct value of the Schrödinger equation in the two universes theirs physical time variables must be reversely related. Thus, the particles that propagate in one of the universes are naturally  identified with the antiparticles that are left in the partner universe. In Sect. \ref{sec4}, we summarise and make some conclusions.

\section{Time reversal symmetry in classical cosmology}\label{sec2}

Let us consider a homogeneous and isotropic spacetime and a scalar field $\varphi$, which represents the matter content of the universe, that propagates minimally coupled to gravity under the action of the potential, $V(\varphi)$. The spacetime is foliated then in homogeneous and isotropic slices, with a total line element given by
\be
ds^2 = - N^2(t) dt^2 + a^2(t) d\Omega_3^2 ,
\ee
where $a(t)$ is the scale factor and $N(t)$ is the lapse function that parametrises the time variable ($N=1$ corresponds to cosmic time), and the homogeneous mode of the scalar field is $\varphi(t)$. Small inhomogeneities around this homogeneous and isotropic background can also be considered \cite{Halliwell1985, Kiefer1987} but as far as the inhomogeneities remain small, the dynamics of the background essentially depends  on the values of the scale factor and the homogeneous mode of the scalar field, $a(t)$ and $\varphi(t)$. From this point of view, the evolution of the universe is determined by the functions $a(t)$ and $\varphi(t)$ that extremize the Hilbert-Einstein action, which for the present case can be written as \cite{Kiefer2007}
\be\label{ACT01}
S = \int dt N \left( \frac{1}{2 N^2} G_{AB} \frac{dq^A}{dt} \frac{dq^B}{dt} - \mathcal V(q) \right) , 
\ee
where, $q^A = \{ a, \varphi \}$, are the coordinates of the configuration space\footnote{For convenience, the initial scalar field $\varphi$ has been rescaled according to $\varphi \rightarrow \frac{1}{\sqrt{2}} \varphi$.},  $G_{AB}$ is given by
\be\label{MSM01}
G_{AB} = {\rm diag}(-a,a^3) ,
\ee
and $\mathcal V(q)$ contains all the potential terms of the spacetime and the scalar field, 
\be\label{POT01}
\mathcal V(q) =  \frac{1}{2} \left( - \kappa a +  a^3 V(\varphi) \right) ,
\ee
where, $\kappa = 0, \pm 1$ for flat, closed and open spatial slices of the whole spacetime. An explicit term for a cosmological constant is implicitly included in the case of a constant value of the potential, $V(\varphi) = \frac{\Lambda}{3}$. The Euler-Lagrange equations derived from the variation of the action (\ref{ACT01}) are \cite{Kiefer2007}
\be\label{FEQ01}
\frac{\ddot a}{a} + \frac{\dot a^2}{2 a^2} + \frac{1}{2 a^2} = - 3 \left(\frac{1}{2} \dot \varphi^2 - V \right)   \,\,\, , \,\,\, \ddot{\varphi} + 3\frac{\dot a}{a} \dot{\varphi} + \frac{\partial V}{\partial \varphi} = 0 .
\ee
The Friedmann equation {\cite{Linde1993}}
\be\label{FEQ02}
\left(\frac{\dot{a}}{a} \right)^2 + \frac{k}{a^2} = \frac{1}{2} \dot{\varphi}^2 + V(\varphi) ,
\ee
turns out to be the Hamiltonian constraint that appears in quantum cosmology from the invariance  of the Hilbert-Einstein action (\ref{ACT01}) under time reparametrisation, $\frac{\delta S}{\delta N}=0$. The field equations (\ref{FEQ01}-\ref{FEQ02}) can in general be difficult to solve analytically  but, anyway, the exact or the approximate solutions of the field equations basically give the evolution of the universe. It is easy to see that these equations are invariant under the reversal change in the time variable, $t \rightarrow -t$. It means that for any given solution $a(t)$ and $\varphi(t)$ one may also consider the symmetric solution, $a(-t)$ and $\varphi(-t)$.

\begin{figure}
\centering
\includegraphics[width=6 cm]{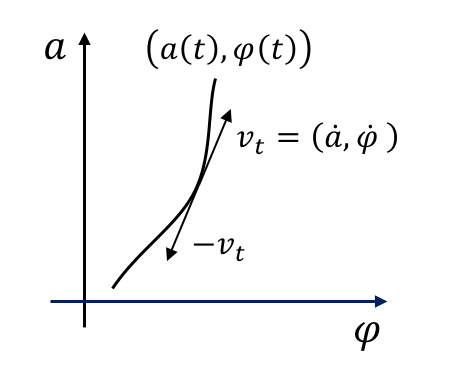}
\caption{The evolution of the universe can be seen as a parametrised trajectory in the minisuperspace, whose parametric coordinates are given by the solutions of the field equation, $a(t)$ and $\varphi(t)$.}
\label{figure01}
\end{figure}

\begin{figure}
\centering
\includegraphics[width=14 cm]{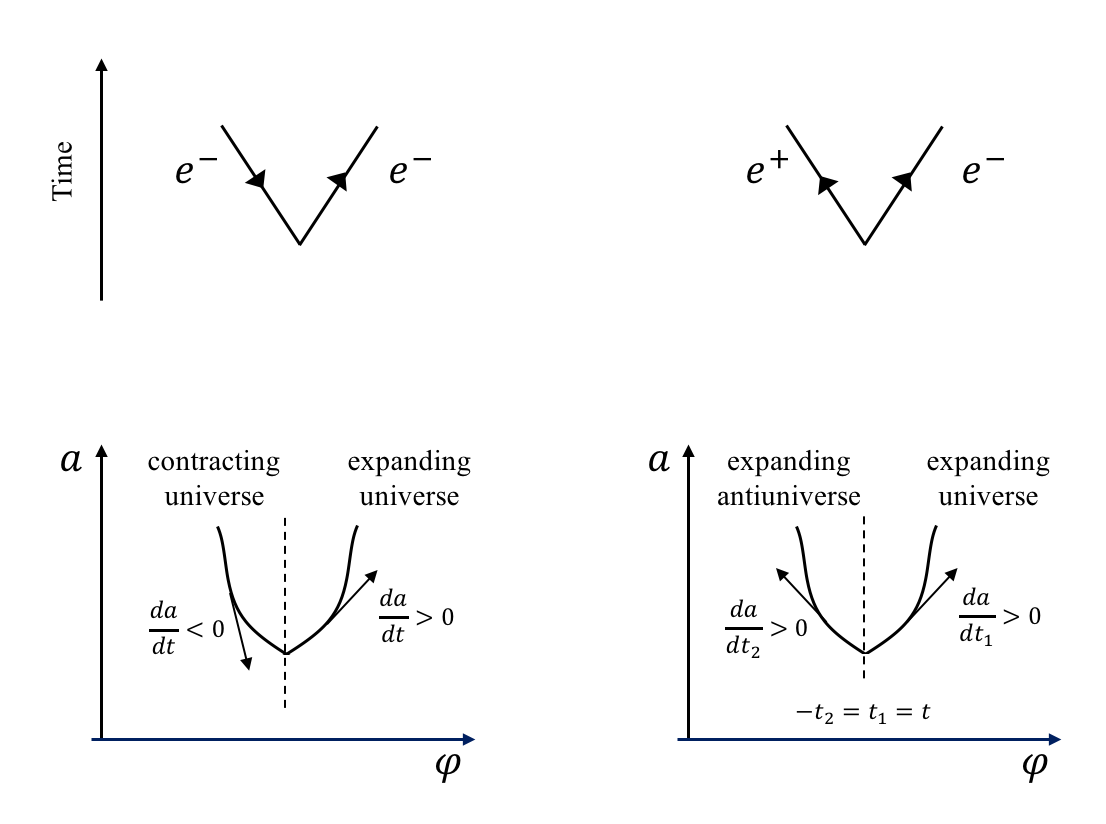}
\caption{Up: the creation of an electron-positron pair (Right) can equivalently be seen as the trajectory of an electron bouncing from backwards to forwards in time (Left). Down: two symmetric cosmological solutions can represent, in terms of the same time variable $t$, a contracting and an expanding universe (Left). In terms of the reversely related time variables, $t_1=-t_2$, the two symmetric solutions represent two expanding universes (Right).}
\label{figure02}
\end{figure}

The action (\ref{ACT01}) and the minisupermetric (\ref{MSM01}) clearly reveal the geometric character of the configuration space, which is called the minisuperspace\footnote{Generally speaking we call superspace to the space of all possible geometries, modulo diffeomorphisms, and all the matter field configurations that can be fitted in those spacetime \cite{Wiltshire2003, Kiefer2007}. However, when we restrict the degrees of freedom by the assumption of some symmetries, like the homogeneity and isotropy that we are considering here, then, it is called the minisuperspace.}, where the scale factor would formally play the role of the time like variable and the scalar field would formally play the role of the spatial like variable\footnote{Let us recall, however, that this is just a formal analogy, and let us also notice that in the case of considering $n$ scalar field minimally coupled to gravity, then, the line element of the minisuperspace would be, $$d \mathfrak s^2 = - a da^2 + a^3 \delta_{ij} d\varphi^i d\varphi^j , $$so the scalar fields would parallel the role of $n$ spatial variables in a $n+1$ dimensional spacetime.}. Therefore, an alternative but equivalent point of view for the evolution of the universe is considering that the time dependent solutions of the scale factor and the scalar field, $a(t)$ and $\varphi(t)$, are the parametric equations of a trajectory in the  minisuperspace, where the time variable {$t$} acts as the (non-affine) parameter in terms of which it is described the trajectory of a 'test universe' (see, Fig. \ref{figure01}). The Euler-Lagrange equations associated to the action (\ref{ACT01}), given by (\ref{FEQ01}), can  be rewritten as the equations of the non-geodesic curves
\be\label{GEO01}
\ddot q^A + \Gamma^A_{BC} \dot q^B \dot q^C = - G^{A B} \frac{\partial \mathcal V(q)}{\partial q^B} ,
\ee
where $G^{AB}$ is the inverse of the minisupermetric $G_{AB}$. The momentum conjugated to the minisuperspace variables can be directly obtained from  (\ref{ACT01}),
\be\label{MOM01}
p_a = -\frac{a \dot a}{N} \,\,\, , \,\,\, p_\varphi = \frac{a^3 \dot \varphi}{N} ,
\ee
and the Hamiltonian constraint, $\frac{\delta H}{\delta N} = 0$,  then reads
\be\label{HC01}
G^{AB} p_A p_B + m^2_{\rm eff}(q) = 0 ,
\ee
where for convenience we have defined\footnote{Written in this way, the resemblance between the description of the trajectory in the minisuperspace and the description of a trajectory in the spacetime is quite evident.}, $m^2_{\rm eff}(q) = 2 V(q)$. In the present case, it yields
\be\label{HC02}
-\frac{1}{a} p_a^2 + \frac{1}{a^3} p_\varphi^2 + m^2_{\rm eff}(q) = 0 ,
\ee
which is the Friedmann equation (\ref{FEQ02}) expressed in terms of the momenta instead of in terms of the time derivatives of the minisuperspace variables. As pointed out above, the geodesic equation (\ref{GEO01}) and the momentum constraint (\ref{HC01}-\ref{HC02}) are invariant under a reversal change of the time variable. It means that the solutions {may} come in pairs with opposite values of the associated momenta (let us notice that the momenta given in (\ref{MOM01}) are not invariant under the same change). From (\ref{MOM01}) and (\ref{HC02}), it is easy to see that in terms of the cosmological time ($N=1$) the two symmetric solutions are
\be\label{FE03}
a \frac{da}{dt} = - p_a = \pm \sqrt{\frac{1}{a^2} p_\varphi^2 + a m^2_{\rm eff}(q) } .
\ee
It clearly reminds to the solutions of the trajectory of a test particle moving in the spacetime \cite{Garay2018}. For instance, in Minkowski spacetime\footnote{A similar procedure can be followed in a curved spacetime.}, the time component of the geodesics satisfies
\be\label{MST01}
\frac{dt}{d\tau} = - p_t = \pm \sqrt{\vec p^2+m^2} ,
\ee
where $\tau$ is an affine parameter and, $p_t = \pm E$, is the energy of the test particle. In the spacetime, the two signs in (\ref{MST01}) represent the opposite values of the time component of the tangent vector to the geodesic, i.e. the two ways in which the geodesic can be run: forward in time and backward in time. This was used by Feynman \cite{Feynman1949} to interpret the trajectories of an electron and a positron as the trajectory of one single electron bouncing from backwards to forwards in time (Fig. \ref{figure02} Up).

In the case of the universe the two solutions given in (\ref{FE03}) also represent two universes: one universe moving forward in the scale factor component and the other moving backward in the scale factor component. In the minisuperspace, however, moving forward in the scale factor component entails an increasing value of the scale factor so  the associated solution represents an expanding universe, and moving backward in the scale factor component entails a decreasing value of the scale factor so the symmetric solution represents a contracting universe. Therefore, the two symmetric solutions form an expanding-contracting pair of universes (see, Fig. \ref{figure02} Down-Left). The total momentum conjugated to minisuperspace variables is conserved because the values of the momenta associated to the two symmetric solutions are reversely related. Let us notice however that the field equations (\ref{FEQ01}) and the Friedmann equation (\ref{FEQ02}) are invariant under a reversal change in the time variable, $t \rightarrow - t$, so from a theoretical point of view we could have chosen $-t$ as the time variable and, then, the solutions that represent an expanding and a contracting universe would have been interchanged. In this paper we are interested in the creation of the universe from the spacetime foam \cite{Hawking1978}, so we shall interpret a contracting-expanding pair of symmetric solutions as the trajectories in the minisuperspace of two newborn universes, both expanding in terms of their reversely related time variables, $t_1=-t_2$ (Fig. \ref{figure02} Down-Right).

Therefore, we shall assume that the universes are created in pairs, both expanding in terms of their internal, reversely related time variables. In terms of the same time variable, however, one of the universes {is} an expanding universe and the other {is} a contracting universe. For instance, for an inhabitant of one of the universes, say Alice, her universe is the expanding one and the partner universe (that she does not see) would be the contracting one. However, it is not contracting for Bob, an inhabitant of the partner universe, for whom things are the other way around, it is his universe the one that is expanding (in terms of his time variable) and Alice's universe, from his point of view, the one that is contracting. Thus, the particles that move in the two universes look like they were propagating backward and forward in time, depending on the observer's point of view. Assuming the CPT theorem, the particles that propagate in the disconnected pieces of the spacetime have consequently opposite values of their charge and parity, so they can be identified in the quantum theory of the composite system with particles and antiparticles. The inhabitants of the two universes can only see the particles that propagate in their own spacetimes but if they would find any signature of the existence of a time reversely related universe, then, they could infer that at the onset primordial antimatter was mainly created in the partner universe, and thus they could conclude that the matter-antimatter asymmetry that they observe in their universes is only an apparent asymmetry that becomes restored in the composite picture of the two symmetric universes.

\section{Quantum cosmology and the creation of universes}\label{sec3}

One would expect that the creation of the universe should have a quantum origin. Therefore, let us analyse the creation of a pair of time reversely related universes in quantum cosmology. The quantum state of the universe is described by a wave function that depends on the metric components of the spacetime as well as on the degrees of freedom of the matter fields. It is the solution of the Wheeler-DeWitt equation, which is essentially the canonically quantised version of the classical Hamiltonian constraint. This is in general a very complicated function. However, as was pointed out in Sec. \ref{sec1}, we are assuming that the universe emerges from the gravitational vacuum and starts inflating from an initial spatial hypersurface $\Sigma(a_i)$ that is small but large enough to consider that the fluctuations of the spacetime are subdominant. In that case, we can model the universe as weakly coupled fields propagating in an essentially homogeneous and isotropic background spacetime. Then, the Hamiltonian of the whole universe can be split into the Hamiltonian of the background and the Hamiltonian of the matter fields \cite{Halliwell1985, Kiefer1987}
\be\label{HC03}
 (\hat{H}_{bg} + \hat{H}_m) \phi = 0 ,
\ee
where the Hamiltonian of the background spacetime, $H_{bg}$, is given by the quantum version of the classical Hamiltonian (\ref{HC02})
\be\label{H0}
\hat{H}_{bg} =  \frac{1}{2 a} \left( {\hbar^2}\frac{\partial^2}{\partial a^2} + \frac{{\hbar^2}}{a} \frac{\partial}{\partial a} - \frac{{\hbar^2}}{a^2} \frac{\partial^2}{\partial \varphi^2} + a^4 V(\varphi) - a^2  \right)  ,
\ee
and $H_m$ contains the Hamiltonian of all the matter fields and their interactions. The wave function, $\phi = \phi(a, \varphi; x_{m})$, where $x_m$ are the variables of the local matter fields, can then be expressed in the semiclassical regime as a linear combination of WKB solutions, i.e. \cite{Hartle1990, Kiefer1992}
\be\label{SCWF01}
\phi = {\sum \phi_+ + \phi_- = } \sum C e^{ \frac{i}{\hbar} S} \psi_+ + C e^{ -\frac{i}{\hbar} S} \psi_-,
\ee
where $C=C(a,\varphi)$ is a real slow-varying function of the background variables, $S=S(a,\varphi)$ is the action of the background spacetime,  $\psi=\psi(a,\varphi; x_{m})$ is a complex wave function that contains all the dependence on the matter degrees of freedom, with $\psi_- = \psi_+^*$, and the sum in (\ref{SCWF01}) extends to all possible classical configurations. A relevant feature to be noticed here is that, because the Hermitian character of the Hamiltonian (\ref{HC03}), which in turn is rooted in the time reversal symmetry of the classical Hamiltonian constraint (\ref{HC02}), the general solution of the Wheeler-DeWitt equation can always be expressed in terms of the two complex conjugated, independent solutions that correspond to the two possible signs in the exponentials of (\ref{SCWF01}). We shall now see that these two wave functions represent similar universes with essentially the same evolution of their spacetimes and similar matter fields propagating therein. However, because the momenta conjugated to the scale factor associated to the two complex conjugated solutions in (\ref{SCWF01}) are reversely related, the time variables of their spacetimes are reversely related too.

In order to see how the  wave functions $\phi_+$ and $\phi_-$ of (\ref{SCWF01}) represent a particular universe one can insert them into the Wheeler-DeWitt equation (\ref{HC03}). After some decoherence process between the two wave functions, which is guarantee for the smallness of the fluctuations of the spacetime \cite{Halliwell1989, Kiefer1992}, one can expect that the Wheeler-DeWitt equation must be satisfied order by order in an expansion in $\hbar$. At order $\hbar^0$ one obtains the following Hamilton-Jacobi equation \cite{Kiefer1987, RP2018a}
\be\label{HJ01}
-\left( \frac{\partial S}{\partial a} \right)^2 +\frac{1}{a^2} \left( \frac{\partial S}{\partial \varphi} \right)^2 + a^4 V(\varphi) - a^2 = 0 .
\ee
This equation contains the dynamics of the background spacetime. It can be converted into the Friedmann equation by defining the WKB time variable given by \cite{Kiefer1987} 
\be\label{WKBt01}
\frac{\partial}{\partial t} = \pm\nabla S \cdot \nabla \equiv \pm\left( -\frac{1}{a} \frac{\partial S}{\partial a}\frac{\partial }{\partial a} +\frac{1}{a^3} \frac{\partial S}{\partial \varphi}\frac{\partial }{\partial \varphi} \right) ,
\ee
where $\nabla$ is the gradient of the minisuperspace \cite{Kiefer1987}. In terms of the WKB time variable,
\be\label{MOM02}
\dot{a}^2 = \frac{1}{a^2} \left( \frac{\partial S}{\partial a}\right)^2 \ , \ \dot{\varphi}^2 = \frac{1}{a^6} \left( \frac{\partial S}{\partial \varphi} \right)^2 ,
\ee
so that the Hamilton-Jacobi equation (\ref{HJ01}) turns out to be the Friedmann equation (\ref{FEQ02}). It thus describes the evolution of the background spacetime.  Furthermore, let us notice that at order $\hbar^0$ the momentum conjugated to the scale factor associated to the wave functions $\phi_+$ and $\phi_-$ is given by
\be\label{MOM03}
p_a = -i\hbar \frac{\partial \phi_\pm}{\partial a} = \pm \frac{\partial S}{\partial a} ,
\ee
where the plus sign corresponds to $\phi_+$ and the minus sign to $\phi_-$. They are thus reversely related and the total momentum associated to the creation of the two universes represented by $\phi_+$ and $\phi_-$  is zero.

\begin{figure}
\centering
\includegraphics[width=14 cm]{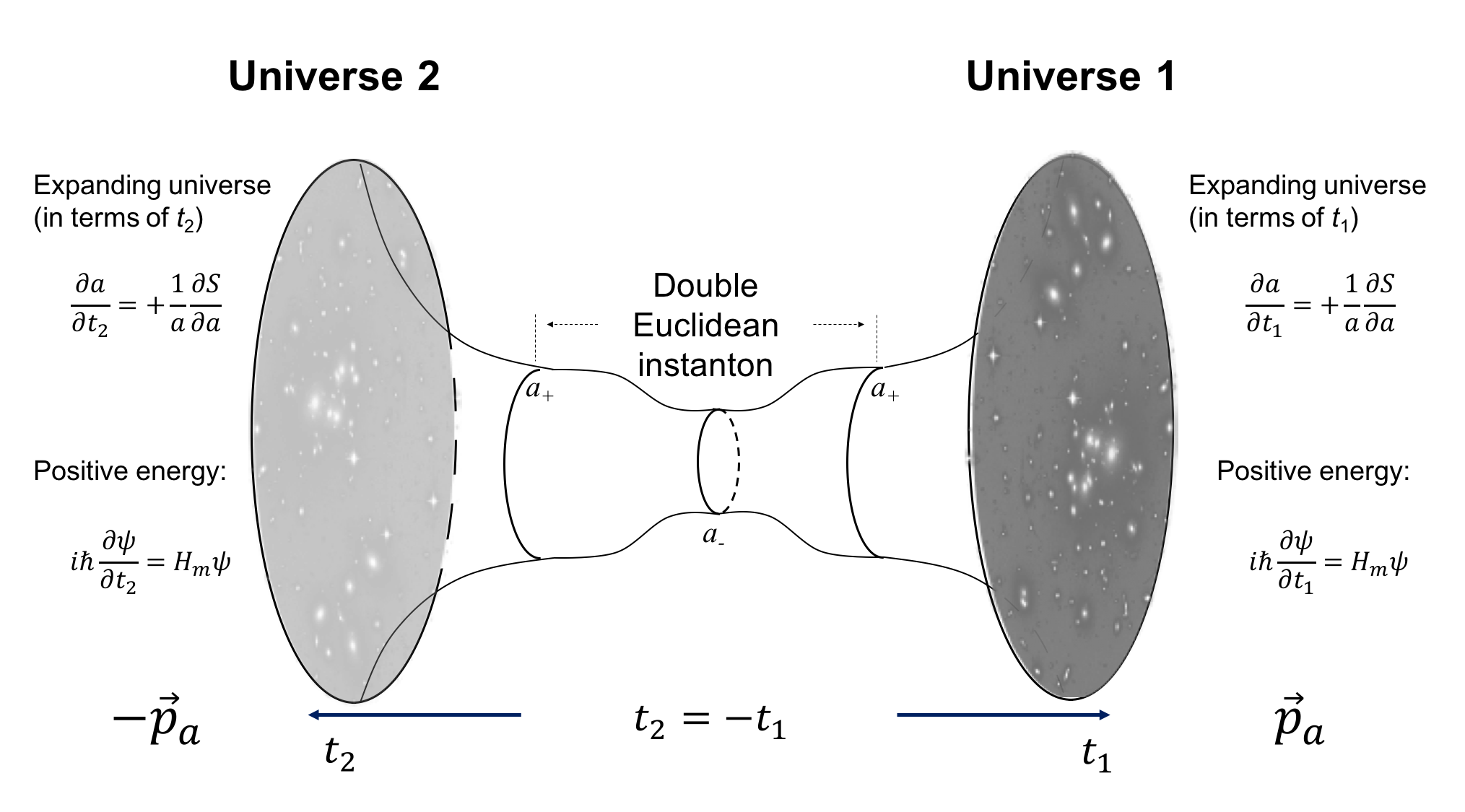}
\caption{The creation of universes in entangled pairs \cite{RP2018a}. The time variables of the two symmetric universes are reversely related. It provides us with the correct value of the Schrödinger equation in the two universes. At the onset, primordial matter would be created in the observer's universe and antimatter in the symmetric one. Particles and antiparticles do not collapse because the Euclidean gap that exists between the two newborn universes \cite{RP2018a, RP2017e}.}
\label{figure03}
\end{figure}

Furthermore, at first order in $\hbar$ in the expansion of the Wheeler-DeWitt equation, one obtains \cite{Kiefer1987, RP2018a} 
\be\label{SCH00}
\mp i \hbar \left( -\frac{1}{a} \frac{\partial S}{\partial a}\frac{\partial }{\partial a} +\frac{1}{a^3} \frac{\partial S}{\partial \varphi}\frac{\partial }{\partial \varphi} \right) \psi = H_m \psi ,
\ee
where the positive and negative signs correspond to $\phi_-$ and $\phi_+$ (\ref{SCWF01}), respectively. The term in brackets in (\ref{SCH00}) is actually the WKB time variable defined in the background spacetime, given by (\ref{WKBt01}), so it means that  (\ref{SCH00})  is essentially the Schrödinger equation for the matter fields that propagate in the classical background spacetime represented by $\phi_+$ and $\phi_-$. We then recover the semiclassical picture of quantum matter fields propagating in a classical background. However, in order to have the proper sign in the Schrödinger equation in each single universe we need to choose the positive sign in the definition of the time variable $t$ in (\ref{WKBt01}) for the wave function  {$\phi_-$} and the negative sign of the time variable for the wave function {$\phi_+$}.  If we assume that the time variable involved in the Schrödinger equation  is the \emph{physical} time variable in the sense that it is the time measured by the observers in their particle physics experiments, so it is the time variable measured by actual clocks, which are eventually made of matter, then, the physical time variables of the two universes are reversely related. It is worth noticing that the eventual inhabitants of the universes will only see the matter of their respective universes and therefore cannot observe the antimatter (from their point of view) that propagates in the symmetric universe. There is also an Euclidean gap between the two universes that prevents matter and antimatter from collapse (see Fig. \ref{figure03}). Therefore, from the point of view of an individual observer there is nothing in principle that makes him suspect the existence of the another universe except perhaps the occurrence of an asymmetry that is hard to explain within the single universe scenario. In principle, it is only from a symmetry consideration that the observer can pose the existence of another universe that justifies the apparent primordial asymmetry between matter and antimatter\footnote{That does not exclude the possibility that other mechanisms of matter-antimatter asymmetry given in the context of a single universe can contribute as well to the total asymmetry observed, but it does assume that the main mechanism would be the creation of matter and antimatter in two separated symmetric universes, provided that the former are not yet fully satisfactory within the Standard Model of particle physics.}.

The creation of particles and antiparticles would follow a similar procedure to that customary considered in the context of a single universe (see, for instance, Ref. \cite{Mukhanov2008}). After the slow roll regime the inflaton field eventually approaches a minimum of the potential and starts oscillating. In each oscillation it decays through different channels and in subsequent stages into the particles of the Standard Model. The interaction with the inflaton field  produces a series of consecutive, non adiabatic change of the vacuum of the matter fields. The associated particle production can then be derived from the Bogolyubov transformation that relates the vacuum states before and after the non adiabatic change, with a particle production given by
\be\label{VS01}
|0 \bar 0 \rangle =\prod_k \frac{1}{|\alpha_k|^\frac{1}{2}}  \left( \sum_n \left( \frac{\beta_k}{2 \alpha_k} \right)^n | n_k, \bar n_{-k} \rangle   \right) ,
\ee
where $|0 \bar 0 \rangle$ is the vacuum state of the corresponding matter field before the oscillation and $| n_k, \bar n_{-k} \rangle$ is the state representing the number of particles and antiparticles created with momentum $k$ after the interaction. The functions $\alpha_k$ and $\beta_k$ are the coefficients of the Bogolyubov transformation that relates the wave functions of the modes before and after the interaction and they contain the effect of the interaction between the inflaton and the matter fields.  In the single universe scenario, because the symmetries of the background, particles and antiparticles are created in perfectly correlated pairs $| n_k, \bar n_{-k} \rangle$ so the number of them is exactly balanced, at least in principle. Different mechanisms that would produce some asymmetry in the creation of matter and antimatter are invoked \cite{}. However, they usually consider some modification or extension of the Standard Model of particle physics.

In the scenario of the twin universes with reversely related time variables presented here, the state of the matter field would still be given by (\ref{VS01}). However, matter and antimatter would be created in different universes so that (\ref{VS01}) should be rewritten as
\be\label{VS02}
|0^{(1)} 0^{(2)} \rangle =\prod_k \frac{1}{|\alpha_k|^\frac{1}{2}}  \left( \sum_n \left( \frac{\beta_k}{2 \alpha_k} \right)^n | n_k^{(1)}, n_{-k}^{(2)} \rangle   \right) ,
\ee
where $n_k^{(1)}$ are the particles in the mode $k$ created in  one of the universes, labeled as $1$, and $n_k^{(2)}$ are the particles in the mode $k$ created in the other universe, labeled as $2$. For the inhabitants of the two universes the primordial antimatter is essentially created in the  partner universe, i.e. for an observer in the universe $1$, $n_k^{(2)}$ are the antiparticles that she misses in her universe and, analogously, $n_k^{(1)}$ are the antiparticles in the mode $k$ that are left from the point of view of an observer of the universe $2$. Particles and antiparticles cannot interact, and therefore cannot annihilate each other, because the Euclidean gap (a quantum barrier) that separates the universes (see Fig. \ref{figure03} and Refs. \cite{RP2018a, RP2017e}).

Let us now make a comment about the influence that the fluctuations of the spacetime could have in the global picture of the creation of the universes in symmetric pairs and the subsequent creation of matter and antimatter separately. The assumption of homogeneity and isotropy made from the very beginning is determining for the time reversal symmetry of the field equations and eventually, for the creation of the universes in symmetric pairs. It means that the scenario presented here is at most a plausible scenario. However, one would expect that as long as the deviation from the homogeneous and isotropic background is relatively small the picture would essentially be rather similar. The small fluctuations of the spacetime would cause the symmetry to stop being an exact symmetry to become an approximate symmetry. It would be expected that the creation of particles and antiparticles cease to occur in perfectly correlated pairs and that (\ref{VS02}) would only be approximate, with corrections at different levels. However, one might expect that this mechanism of creating matter and antimatter separately could still be dominant, as long as the deviations from the homogeneous and isotropic spacetime were small. In addition, other mechanisms that produce matter and antimatter asymmetry can still be considered and they could work jointly to produce the observed amount of asymmetry. Perhaps, the creation of universes in (non perfectly) symmetric pairs could relax the tensions or the anomalous fine tuning that might exist between the theoretical models and the observational data. It might also entail other observable effects \cite{Boyle2018}. Therefore, we consider that it is an appealing scenario that deserves a further analysis and a deeper understanding.

\section{Conclusions}\label{sec4}

The evolution of the universe can be seen as a worldline of the minisuperspace formed by the scale factor, which formally plays the role of a time-like variable, and the scalar fields, which formally play the role of the spatial components. From that point of view, the time reversal symmetry of the field equations becomes equivalent to the invariance of the geodesics of the minisuperspace under a reversal parametrisation of their non-affine parameter.

Positive oriented paths with respect to the scale factor component in the minisuperspace entail an increasing value of the scale factor so they represent expanding universes. On the contrary, negative oriented worldlines with respect to the scale factor component represent contracting universes. However, because the time reversal invariance of the Lagrangian of a homogeneous and isotropic spacetime, the terms 'expansion' and 'contraction' can be interchanged so we end up with four symmetric solutions, two  oriented forward and two oriented backwards. The former represent two expanding universes with their time variables reversely related and the latter represent two virtual universes that rapidly return to the gravitational vacuum from which they emerged.

Quantum mechanically, the solutions of the Wheeler-DeWitt equation can always be given in complex conjugated pairs because the Hermitian character of the Hamiltonian constraint. Two complex conjugated solutions entail opposite values of the momentum conjugated to their scale factors so the creation of universes in pairs whose wave functions are complex conjugated has associated a total zero momentum. Furthermore, the analysis of the emergence of the semiclassical spacetime in the universes that they represent suggests that the physical time variables of their spacetimes should be reversely related. Then, the inhabitants of the two universes would naturally identify matter with the particles that propagate in their spacetimes and the unobserved primordial antimatter with the particles that would propagate in the symmetric universe.

The consideration of a homogeneous and isotropic spacetime makes the scenario presented here to be at most a plausible one. However, it might be realistic if our universe was created and started inflating from an initial hypersurface that is small but large enough to assume that the fluctuations of the spacetime are subdominant. In that case, small deviations from the homogeneous and isotropy of the spacetime would transform the exact time reversal symmetry into an approximate one. Nevertheless, as long as the deviations are relatively small, one would expect that the global picture would not differ so much from the one depicted here. The creation of matter and antimatter in separated universes might still be the dominant one, or at least it could help or enhance other mechanisms that are already considered in the context of a single scenario.

Finally, this work sets the arena for a deeper and more detailed study, which might eventually unveil whether the universe is actually part of a universe-antiuniverse pair.



\bibliographystyle{apsrev4-1}
\bibliography{../bibliography}

\end{document}